\documentclass[fleqn,usenatbib]{mnras}

\usepackage{newtxtext,newtxmath}

\usepackage[T1]{fontenc}
\usepackage{ae,aecompl}

\usepackage{graphicx}	
\usepackage{dblfloatfix}
\usepackage{amsmath}	
\usepackage{threeparttable}
\usepackage{subcaption}
\usepackage{caption}
\usepackage{color}
\usepackage[dvipsnames]{xcolor}
\usepackage[normalem]{ulem}
\usepackage{booktabs}

\usepackage{multirow}
\def \link_col{blue}

\title[Gamma-ray observation towards M17]{Gamma-ray observation towards the young massive star cluster NGC\,6618 in the M17 region}

\author[ et al.]{
Bing Liu,$^{1,2,3}$
Rui-zhi Yang,$^{1,2,3}$\thanks{E-mail: yangrz@ustc.edu.cn}
Zhiwei Chen$^4$
\\
$^{1}$Department of Astronomy, School of Physical Sciences, University of Science and Technology of China, Hefei, Anhui 230026, China\\
$^{2}$CAS Key Labrotory for Research in Galaxies and Cosmology, University of Science and Technology of China, Hefei, Anhui 230026, China \\
$^{3}$School of Astronomy and Space Science, University of Science and Technology of China, Hefei, Anhui 230026, China \\
$^4$Purple Mountain Observatory, Chinese Academy of Sciences, Yuanhua Road 10, 210023 Nanjing, China
}

\date{Accepted XXX. Received YYY; in original form ZZZ}

\pubyear{2021}
\begin{document}
\label{firstpage}
\pagerange{\pageref{firstpage}--\pageref{lastpage}}
\maketitle

\begin{abstract}

Young massive clusters have been established as a new population of gamma-ray sources and potential cosmic ray (CR) accelerators. In this paper, we report the detection of  gamma-ray emissions near the young star cluster NGC 6618, which is one of the youngest star clusters in our Galaxy. The detected gamma-ray emissions can be divided into two components. One component is point-like and reveals harder spectrum,  while the other is extended and with softer spectrum.  Such spectral features are significantly different from other young massive clusters and  may be due to the propagation effects of CRs accelerated in NGC 6618. 
\end{abstract}


\section{Introduction}

Young massive star clusters (YMCs) are established as a new population of gamma-ray sources in recent years and are believed to be able to account for the acceleration of a significant part of Galactic cosmic rays (CRs) \citep{aharonian19}. Several such systems are detected in GeV or TeV gamma-ray band, with significant spatial extensions up to more than 50\,pc and a universal hard gamma-ray spectrum which can be described by a powerlaw function with an index of $2.2-2.4$. More interestingly, the derived CR spatial distribution reveals a $\frac{1}{r}$ profile, which is consistent with a continuous injection of CRs. Considering the size of the source, the CRs are believed to be injected in the time scale of  $\ga 10^5$ years, which is longer than the acceleration phase of supernova remnants (SNRs). Moreover, the massive star winds, which can be as high as $\ga 3000~\rm km\, s^{-1}$ in O-type stars and Wolf-Rayet stars, are powerful enough to accelerate CRs to high energies nearly in the whole lifetime of the massive stars \citep{Cesarsky83}. In this regard, the CRs in YMCs are more likely accelerated by "live" stars rather than the "dead" stars. In an observational point of view, several YMCs have been detected in gamma-rays, such as Cygnus Cocoon\citep{Ackermann11}, NGC 3603 \citep{yang17,ngc3603_20}, Westerlund 1 \citep{hess_wd1}, Westerlund 2 \citep{yang18}, RSGC 1\citep{Katsuta17, sun20b}, W40 \citep{sun20a} and W43 \citep{yang20w43}.

The YMC NGC 6618, with an age less than 3 Myr \citep{2002ApJ...577..245J}, is one of the youngest massive star clusters in our Galaxy, and contains hundreds of stars earlier than B9 including dozens of O stars \citep{1980A&A....91..186C,hoffmeister08}. The OB stars in the cluster, especially the most massive ones such as the O4 binary system \citep[CEN 1a + CEN 1b;][]{hoffmeister08}, provide the ionizing power for the \ion{H}{ii} region, M17 nebula (also known as  Omega Nebula or W38), which is located in the Sagittarius spiral arm \citep{1979ApJ...230..415E,2019ApJ...885..131R} at a distance $\approx$ 2.0\,kpc \citep{2011ApJ...733...25X,2016MNRAS.460.1839C}. Moreover, the cluster is also thought to be the source of energetic feedback to the dense molecular clouds (MCs) surrounding  M17 nebula \citep{Quang2020}.



The MC associated with M17 was first detected in CO molecular line emission at velocity around $20\,\mathrm{km\,s^{-1}}$, and the line emission extends along the Galactic plane for at least 85 pc and has a total mass in the order of $10^6\,{\rm M}_{\sun}$ \citep{1976AJ.....81.1089E}, comparable to the typical definition of the giant MC (GMC). The M17 GMC is usually split into three regions according to the relative location with respect to NGC\,6618. The cloud in the north, i.e., M17 North cloud, has the smallest mass and size \citep{2003ApJ...590..895W}. The adjacent cloud in the southwest, i.e., M17 SW cloud, is the densest region and highly clumpy \citep{1990ApJ...356..513S, Quang2020}, and contributes most of the forming stars of low- to high-mass \citep{2009ApJ...696.1278P,2012PASJ...64..110C,2013A&A...557A..51C,2015A&A...578A..82C,2021ApJ...922...90C,2017A&A...604A..78R,2020ApJ...888...98L}. The cloud extending to the very southwest, i.e., M17\,SWex, appears as an infrared dark cloud and hosts sites of low- to intermediate-mass star formation \citep{2010ApJ...714L.285P,2016ApJ...825..125P,2022RAA....22c5021Y}. By comparing the dense gas properties of M17\,SW and M17\,SWex, \citet{Quang2020} found that the clumps in M17\,SW are denser, more compact, and more gravitationally bound than those in M17\,SWex.

If the massive stars in NGC\,6618 can accelerate protons to very high energies, then these protons could illuminate the clouds around them via proton-proton inelastic collisions.
Recently released {\it Fermi}-LAT\  10-year Source Catalog  \citep[4FGL-DR2,][]{4fgl_dr2} revealed a gamma-ray point source that located in the direction of M17 region (named as 4FGL J1820.4-1609c), which is possibly associated with NGC\,6618. In such a complex region as described above, the origin of these gamma-rays is still an intriguing puzzle and requires a careful and comprehensive investigation.  In the following section,  we present the details of the analysis of {\it Fermi}-LAT data and the results. Then in Sect.~\ref{sec:gas},  we investigate the gas distributions around M17.  Next in Sect.\ref{sec:dis}, we combine the multiwavelength observations with gamma-ray results to explore the radiation mechanism of the gamma-rays.  Finally, we summarise the main conclusions and implications of our study in Sect.\ref{sec:con}.

\section{{\it Fermi}-LAT data analysis and results}
\label{sec:gray}

To study the gamma-ray emission in M17 region, we collected about 12.5 years (from 2008-08-04 15:43:36 (UTC) to 2021-04-16 23:02:55 (UTC)) of {\it Fermi}-LAT Pass 8 data, and used Fermitools from Conda distribution\footnote{https://github.com/fermi-lat/Fermitools-conda/} and applied the latest version of the instrument response functions {\it P8R3\_SOURCE\_V3}. 
Given the energy-dependent point-spread function (PSF) of the LAT data, we used different datasets to optimize the spatial and spectral analyses.
The region of interest (ROI), centered at M17 (R.A.$=275.195^{\circ}$, Dec.$= -16.172^{\circ}$, J2000) is adjusted for each dataset correspondingly.
The source models, generated by {\sl make4FGLxml.py}\footnote{https://fermi.gsfc.nasa.gov/ssc/data/analysis/user/make4FGLxml.py}, consist of the sources in 4FGL-DR2 within the ROI enlarged by $10^\circ$, the Galactic diffuse background emission (\emph{gll\_iem\_v07.fits}), and isotropic emission (\emph{iso\_P8R3\_SOURCE\_V3\_v1.txt}).

\subsection{Spatial analysis and results}
\label{subsec:spatial}
In order to study possible energy dependent morphology in M17 region, we conduct separate spatial analyses for gamma-rays within different energy ranges, 
including 0.5$-$5\,GeV, 5$-$500\,GeV, and 0.5$-$500\,GeV, hereinafter referred as LE, HE and ALL datasets, respectively. Specific parameters of the cuts and details of each dataset are provided in Table.\ref{tab:data}.

\begin{figure*}
\includegraphics[width=0.45\textwidth]{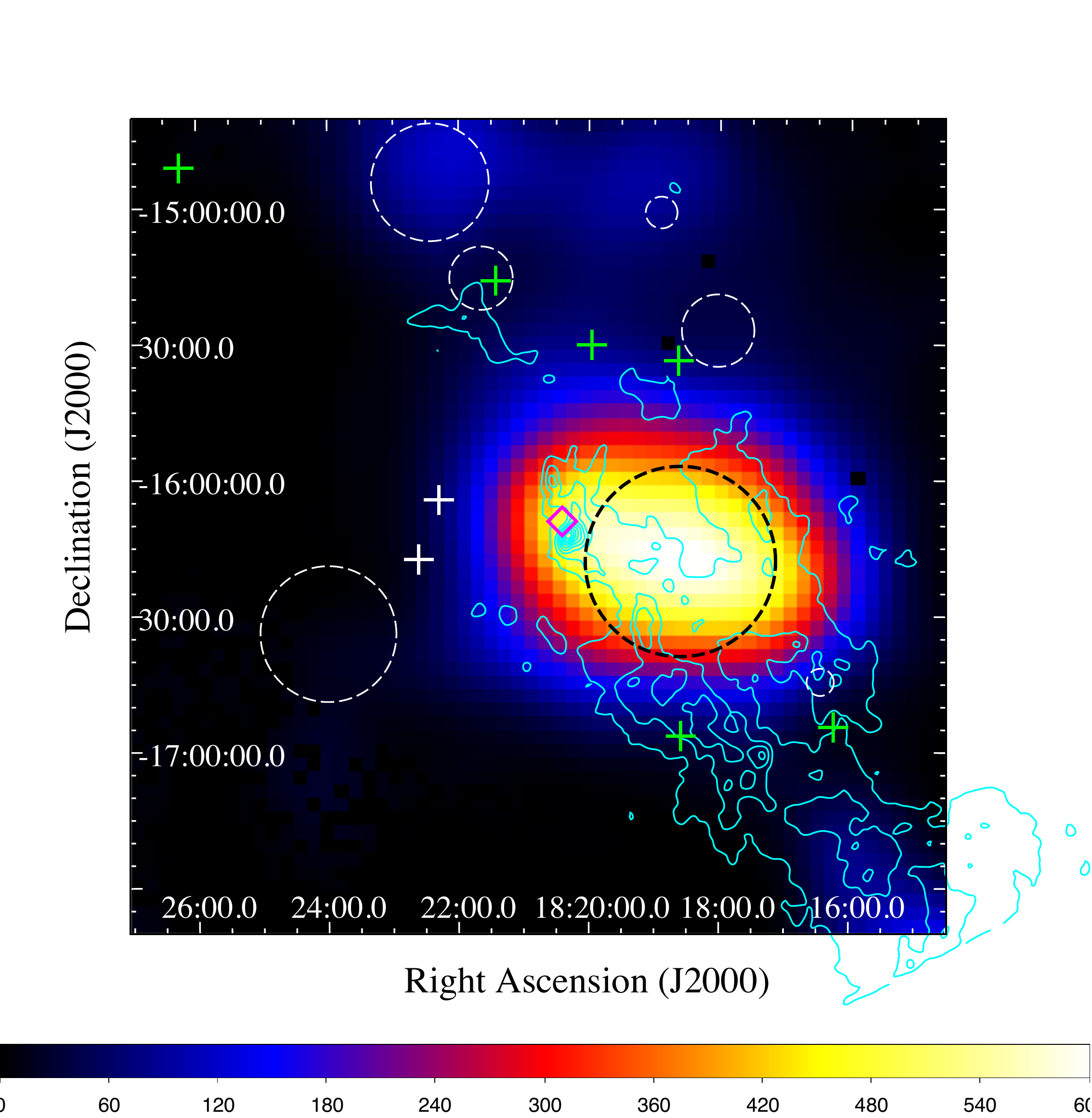}
\includegraphics[width=0.45\textwidth]{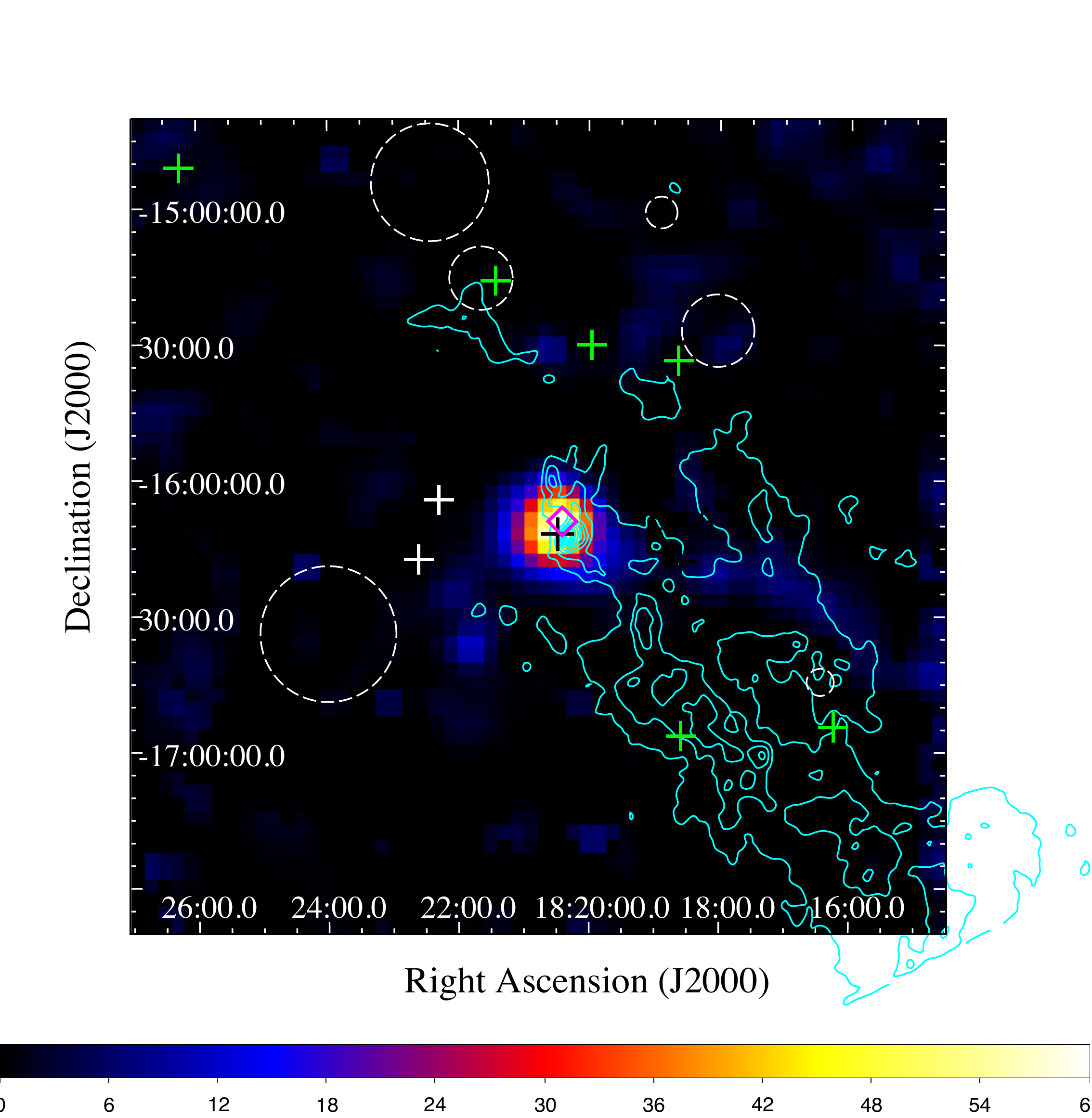}
\caption {Smoothed residual TS maps centered at M17 for LE dataset (left) and  HE dataset (right), respectively. The green crosses show the position of nearby sources listed in 4FGL-DR2, and the magenta diamond represents 4FGL J1820.4-1609c, which is removed from the fitted background model. The white dashed circle and crosses illustrate the nearby supernova remnants and pulsars, respectively.  The black dashed circle illustrate the best-fit gaussian disk model (Src A) from fitting LE dataset, and the black cross represents the position of the point source (Src B) for fitting HE dataset.   The maps are overlaid with contours generated from  $^{13}$CO ($J$=1-0) integrated intensity map in the velocity range 10-34\,$\mathrm{km\,s^{-1}}$  at ten linear scale levels between 15 and 150 $\mathrm{K\,km\,s^{-1}}$.  See Sect.~\ref{subsec:spatial} for details.
}
\label{fig:tsmaps}
\end{figure*}

Following procedures were applied to the LE and HE datasets respectively, in which the spectral type of the tested sources are all set to be simple PowerLaw (PL).
At first, we used {\sl gtlike} to optimize the source models.
Since 4FGL J1820.4-1609c is very likely associated with NGC\,6618 and cannot be treated as background. Thus we removed 4FGL J1820.4-1609c from the optimized background source models,
and generated the residual test statistic (TS) maps.
As shown in Fig.\ref{fig:tsmaps},  the residual gamma-rays of LE dataset are very diffusive and shifted to the west of M17, meanwhile, the gamma-rays of HE dataset are rather point-like and in the direction of NGC\,6618.
Therefore, to find out whether the residual emission is extended or not, we added a point-like source or an extended source (using RadialGaussian model) to the modified source model, from which 4FGL J1820.4-1609c are deleted.  The added point or the center of the gaussian disk is set at the peak position of the residual TS map for each dataset, which is presented in Table~\ref{tab:spatial}. 
The radius ($\sigma_{\rm disk}$) of the disk varies from  $0.1^\circ$ to $0.5^\circ$ with a step of $0.05^\circ$. 
Then we fitted above modified models  to calculate the corresponding $\rm TS_{\rm ext}$, defined as $-2\log({\mathcal L}_{\rm p}/{\mathcal L}_{\rm ext})$. 
The $L_{\rm p}$ is the maximum likelihood of the model with a point source added at the peak position, and the $\mathcal L_{\rm ext}$ is the acquired maximum likelihood of the tested gaussian disk models.  For LE dataset, the maximum $\rm TS_{\rm ext}$ (290.6) is obtained when $\sigma_{\rm disk}$=$0.35^\circ$, meaning the significance of the extension is about 17.0. However, for HE dataset, gaussian disk models show no improvement compared to the point source assumption. The specific results for the spatial analyses of LE and HE datasets are listed in Table~\ref{tab:spatial}. 

Finally, we analysed  ALL dataset in order to test whether one source with the spatial model obtained from LE or HE dataset (1-source  hypothesis) or two sources with the spatial models acquired from both LE and HE datasets (2-source hypothesis) could represents the overall data better. By comparing the  maximum log(likelihood) of these models, we obtained the significance of 2-source hypothesis versus 1-source hypotheses or the non-modified 4FGL-DR2 model is $\sim 9 \sigma$. 

\subsection{Spectral analysis and results}
\label{subsec:spec}
From above analysis, we noticed that the morphology of gamma-ray emission around M17 region is energy-dependent and these gamma-rays are better represented as two separate components. 
In the following spectral analysis applied to ALL dataset, we replace 4FGL J1820.4-1609c with two sources: one source is an extended source with  $\sigma_{\rm disk}=0.35^\circ$  lying to the west of NGC\,6618 (hereafter referred as Src A),  and the other is a point-like source that is coincided with NGC\,6618 (hereafter referred as Src B).

To find out the spectral shape of Src A and Src B,  we performed likelihood-ratio test for spectral models including LogParabola (LogP), PLSuperExpCutoff2 (PLEC), BrokenPowerLaw (BPL), in which the PL model is the null hypothesis. 
The formulae and free parameters of these spectral model are presented in Table.\ref{tab:form}. 
The significance of the tested model $\sigma_{\rm model}$ is defined as $\sqrt{-2\log({\mathcal L}_{\rm PL}/{\mathcal L}_{\rm model})}$.
Firstly, we changed the spectral type of Src A  from PL to  LogP, PLEC and BKL respectively, keeping the spectral model of Src B to be PL.  Next, we set the spectral model of Src A to be PL, and change the spectral type of Src B from PL to LogP, PLEC, and BPL, respectively. 
As shown in Table.\ref{tab:spectral}, those more complicated spectral models do not improve the overall fitting results ($\sigma_{\rm model}<3$), and the simple PL model is capable to  represent their spectral shape.  The best-fit results of the PL model for Src A and Src B are listed in Table.\ref{tab:flux}.
 
\begin{figure*}
\includegraphics[width=0.45\textwidth]{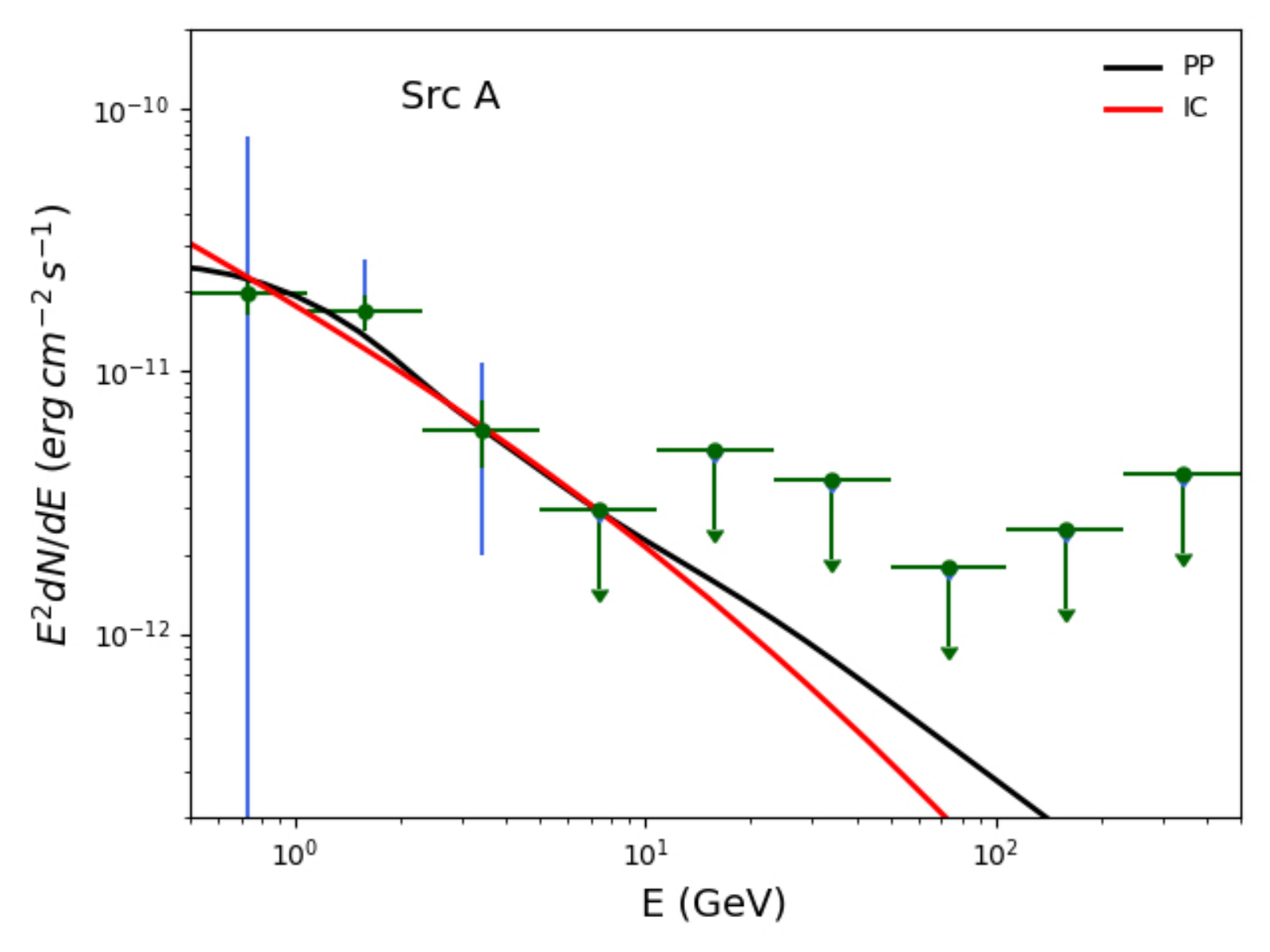}
\includegraphics[width=0.45\textwidth]{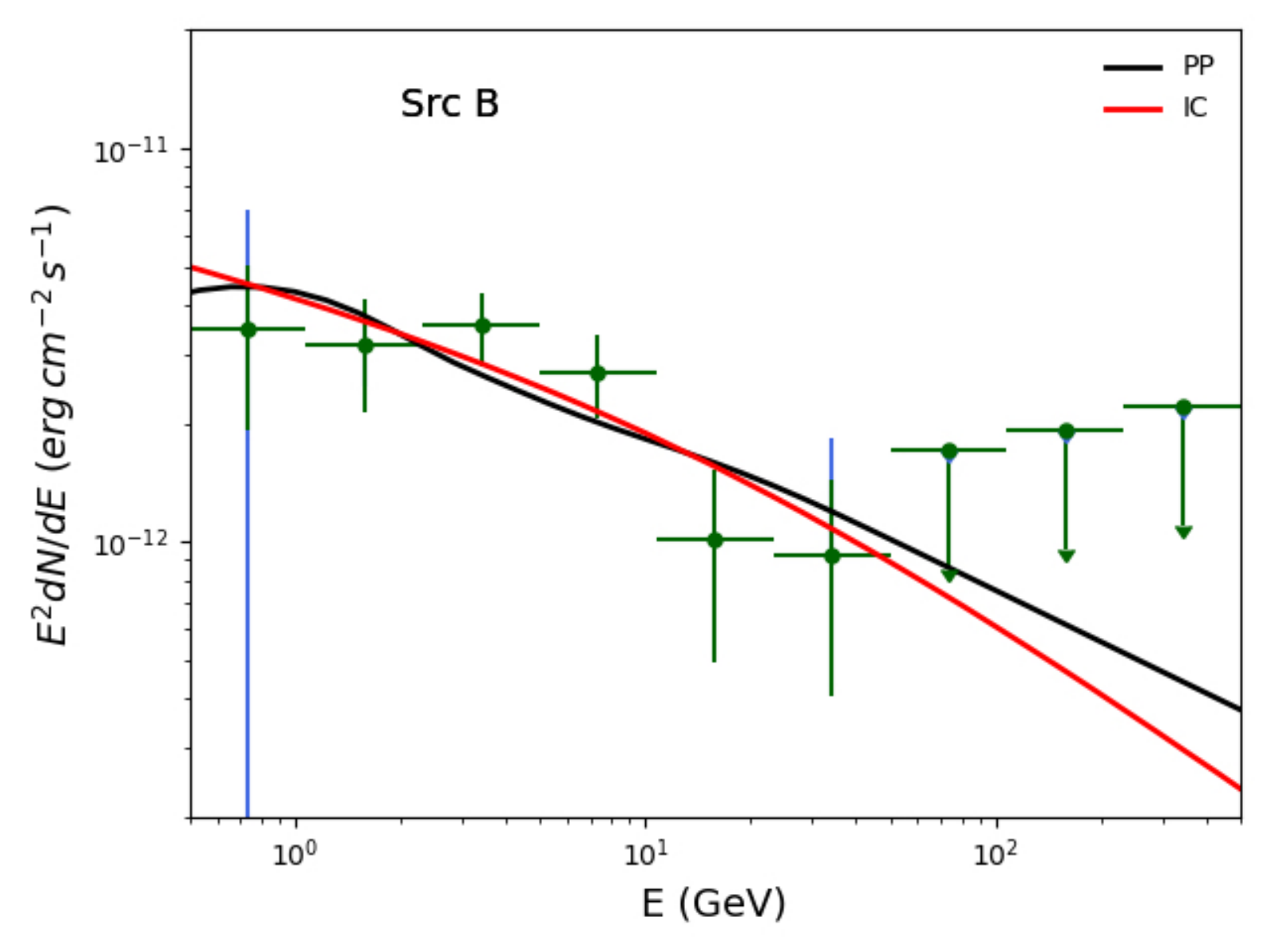}
    \caption { SEDs of Src A (left) and Src B (right) respectively. For each data point, the green error bar indicates the uncertainty caused by the statistics and the effective area, and the blue error bar shows the uncertainty caused by the imperfection model of the Galactic diffuse background. The best-fit PP models (black lines) and IC models (red lines) for Src A and Src B are represented.
    See Sect.~\ref{subsec:spec}  and Sect.~\ref{sec:dis} for details.}
\label{fig:sedfit}
\end{figure*}

The spectral energy distributions (SEDs) of Src A and Src B as shown in Fig.\ref{fig:sedfit} were extracted from the maximum likelihood analysis of source class events in nine logarithmically spaced energy bins within 0.5--500 GeV. Expect the energy range of the selected events, other parameters applied for the data preparation are the same as the ALL datasets. During the fitting process, the free parameters only include the normalization parameters of the sources with the significance $\geq5\sigma$ within $10^{\circ}$ from ROI center as well as the Galactic and isotropic diffuse background components, while all the other parameters are fixed to their best-fit values from above analysis of ALL dataset in which the spectral shape of Src A and Src B are assumed to be PL. In addition to the uncertainties caused by the statistics and the effective area (green error bars in Fig.\ref{fig:sedfit}), we also estimated the uncertainties caused by the imperfection model of the Galactic diffuse background (blue error bars in Fig.\ref{fig:sedfit}) by artificially varying its normalization by $\pm6\%$ from the best-fit value of each energy bin  and recording flux deviation of the source due to above changes as the systematic error, following the method from \citet{Abdo09}.
We note that in lower energy range (0.5--1 GeV, first bin of the SEDs), the fluxes of both Src A and Src B drop dramatically when the Galactic diffuse background was artificially enhanced. 

\begin{table}
\caption{Description of the datasets for spatial analysis}  
\label{tab:data}       
\begin{tabular}{cccc}       
\hline\hline     
 Name & ALL &LE &HE\\
 \hline
E$_{min}$ (GeV) & 0.5 &0.5 & 5 \\
E$_{max}$(GeV) & 500 &5 & 500 \\
ROI & $20^{\circ}\times20^{\circ}$ &  $20^{\circ}\times20^{\circ}$ & $10^{\circ}\times10^{\circ}$ \\
z$_{max}$ & $90^{\circ}$ &$90^{\circ}$  &$105^{\circ}$ \\
evclass & 128 & 128 & 128 \\
evtype &  16/32 & 32 & 16\\
N$_{ebins}$  &30 &10 & 16\\
DATA\_QUAL& $>0$& $>0$& $>0$\\
LAT\_CONFIG&1 & 1& 1\\
free{$^{a}$}/all{$^{b}$} & $10^{\circ}$/ $25^{\circ}$& $10^{\circ}$/ $25^{\circ}$&$5^{\circ}$/ $15^{\circ}$\\    
\hline
\end{tabular}
\\
{\footnotesize  $^{a}$ The radius in degrees from the center of ROI within which the  spectral parameters are free to change for sources with significance $>5$. \\
$^{b}$ The radius in degrees from the center of ROI within which sources are included in the XML file.}
\end{table}

\begin{table}
\caption{Spatial analysis results of the LE and HE datasets}  
\label{tab:spatial}       
\begin{tabular}{ccc}       
\hline\hline     
 Name &LE &HE\\
 \hline
Position (Ra, Dec) &(274.65$^{\circ}$ ,-16.30$^{\circ}$) &(275.120$^{\circ}$ ,-16.20$^{\circ}$)  \\
$\mathcal  L_{\rm 0}$ &645931.3 &-36584.1\\
$\mathcal  L_{\rm p}$  &646128.2 &-36579.1 \\
$\mathcal  L_{\rm ext}$  &646273.5  &-36578.8 \\
$\sigma_{\rm disk}$ ($^{\circ}$ ) &$0.35\pm0.05$ & $0.10\pm0.05$\\
${\rm TS}_{\rm ext}$  &290.6   & 0.3  \\
\hline
\end{tabular}
\\
{\footnotesize The position refers to the peak location of the residual TS map for each dataset. $\sigma_{\rm disk}$ is the radius of gaussian disk of the maximum -log(likelihood) among the extension test models in which $\sigma_{\rm disk}$ varies from  $0.1^\circ$ to $0.5^\circ$ with a step of $0.05^\circ$ and the center of the disk located at the corresponding peak position.
The $\mathcal  L_{\rm 0}$ is the likelihood of the non-modified 4FGL-DR2 model. The $\mathcal  L_{\rm p}$ is the likelihood of the model with 4FGL J1820.4-1609c removed and a point source added at the peak position. 
The $\mathcal  L_{\rm ext}$ is the -log(likelihood) of best-fit gaussian disk model.  ${\rm TS}_{\rm ext}$ =$-2\log({\mathcal L}_{\rm p}/{\mathcal L}_{\rm ext})$.}
\end{table}

\begin{table}
\caption{Formulae for gamma-ray spectral analysis}
\label{tab:form}
\begin{tabular}{lll} 
\hline \hline
Name & Formula  &Free parameters   \\
\hline
PL   & $\frac{\mathrm{d}N}{\mathrm{d}E}$ = $N_0 {\left(\frac{E}{E_0}\right)}^{-\Gamma}$   & $N_0$, $\Gamma$  \\
LogP & $\frac{\mathrm{d}N}{\mathrm{d}E}$ = $N_0 \left(\frac{E}{E_\mathrm{b}}\right)^{-\Gamma - \beta \log(\left(\frac{E}{E_\mathrm{b}}\right)} $  & $N_0$, $\Gamma$, $\beta$\\
PLEC & $\frac{\mathrm{d}N}{\mathrm{d}E}$ = $ N_0 \left(\frac{E}{E_0}\right)^{-\Gamma_1} \exp\left({-\left({\frac{E}{E_\mathrm{cut}}}\right)^{\rm b}}\right)$  &  $N_0$, $\Gamma$, $E_\mathrm{cut}$, {\rm b}\\
BPL &   $\frac{\mathrm{d}N}{\mathrm{d}E}$ = $\begin{cases} N_0{\left(\frac{E}{E_\mathrm{b}}\right)}^{-\Gamma_1} &  \mbox{: } E<E_\mathrm{b} \\ N_0\left(\frac{E}{E_\mathrm{b}}\right)^{-\Gamma_2} & \mbox{: }E>E_\mathrm{b} \end{cases}$ & $N_0$, $\Gamma_1$, $\Gamma_2$, $E_{\rm b}$ \\
\hline
\end{tabular}
\end{table}

\begin{table}
	\caption{ The significance of  the tested spectral models ($\sigma_{\rm model}$)  for Src A and Src B} 
	\label{tab:spectral} 
	\begin{tabular}{c|ccc}
	\hline
	\hline
    Spectral Model    & LogP     & PLEC   & BPL         \\
    \hline
   Src A$^{a}$   & 2.3  &  2.3  & 2.2 \\
   \hline
  Src B$^{b}$   & 2.5   & 2.3   & 2.3  \\
\hline
\end{tabular}
\\
{\footnotesize  $^{a}$ The spectral type of Src B is PL}
\\
{\footnotesize  $^{b}$ The spectral type of Src A is PL}
\end{table}

 
\begin{table}
	\caption{The best-fit results of the PL model for Src A and Src B} 
	\label{tab:flux} 
	\begin{tabular}{c|cccc}
	\hline
	\hline
        & $\Gamma$ &Flux   &  Luminosity  $^{a}$  &  significance $^{b}$    \\
        & &($\,\rm{ph}\,{\rm cm}^{-2}\,{\rm s}^{-1}$)& ($\rm{erg}\,{\rm s}^{-1}$)& $\sigma$ \\
    \hline
   Src A   & $2.73\pm0.08$ &$1.58\times10^{-8} $ & $1.14\times10^{33} $  & 15 \\ 
   \hline
  Src B  &$2.41\pm0.06$  &$5.67\times10^{-9} $   & $5.67\times10^{32} $ & 12  \\ 
\hline
\end{tabular}
\\
{\footnotesize  $^{a}$ The distance  of Src A and Src B is assumed to be 2\,kpc.}
\\
{\footnotesize  $^{b}$ The significance of each source $\simeq \sqrt{\rm TS}$.}
\end{table}

\section{Gas distributions near M17}
\label{sec:gas}
In this work, we analysed the large scale $^{12}$CO/$^{13}$CO ($J$=1-0) molecular line data toward the M17 GMC, as a part of the Milky Way Imaging Scroll Painting (MWISP) project. The details of MWISP project are referred to \citet{2019ApJS..240....9S}. The $^{12}$CO/$^{13}$CO ($J$=1-0) molecular data used in this work span from the galactic longitude $13\degr$ to $16\degr$ and the galactic latitude $-1\fdg5$ to $0\fdg5$. After fitting the baseline and calibrating the main beam efficiency, the reduced 3D data cubes with a grid spacing of $30\arcsec$ have a typical root mean square noise level of $\sim 0.5\,\mathrm{K}$ for $^{12}$CO ($J$=1-0) transition line and $\sim0.3\,\mathrm{K}$ for $^{13}$CO ($J$=1-0) transition line at a channel width of $0.16\,\mathrm{km\,s^{-1}}$. The spatial resolution of the CO data is $\sim50\arcsec$. 

The large-scale $^{12}$CO and $^{13}$CO gas distributions over the area from the galactic longitude $13\fdg67$ to $15\fdg39$ and the galactic latitude $-0\fdg87$ to $-0\fdg27$ were studied earlier in \citet{Quang2020}; however, the spatial extent of Src A is not fully covered by this research. 
The bulk of the M17 GMC emission is seen in the velocity range $10-30\,\mathrm{km\,s^{-1}}$ from the Sagittarius arm, while the components at $\sim37-38\,\mathrm{km\,s^{-1}}$ and $\sim57\,\mathrm{km\,s^{-1}}$ are from the spiral arms at larger distances \citep{2016RAA....16...56Z,Quang2020}. Therefore, we only considered the emission around $10-30\,\mathrm{km\,s^{-1}}$ for the MWISP CO data.

\begin{figure*}
\includegraphics[width=0.9\textwidth]{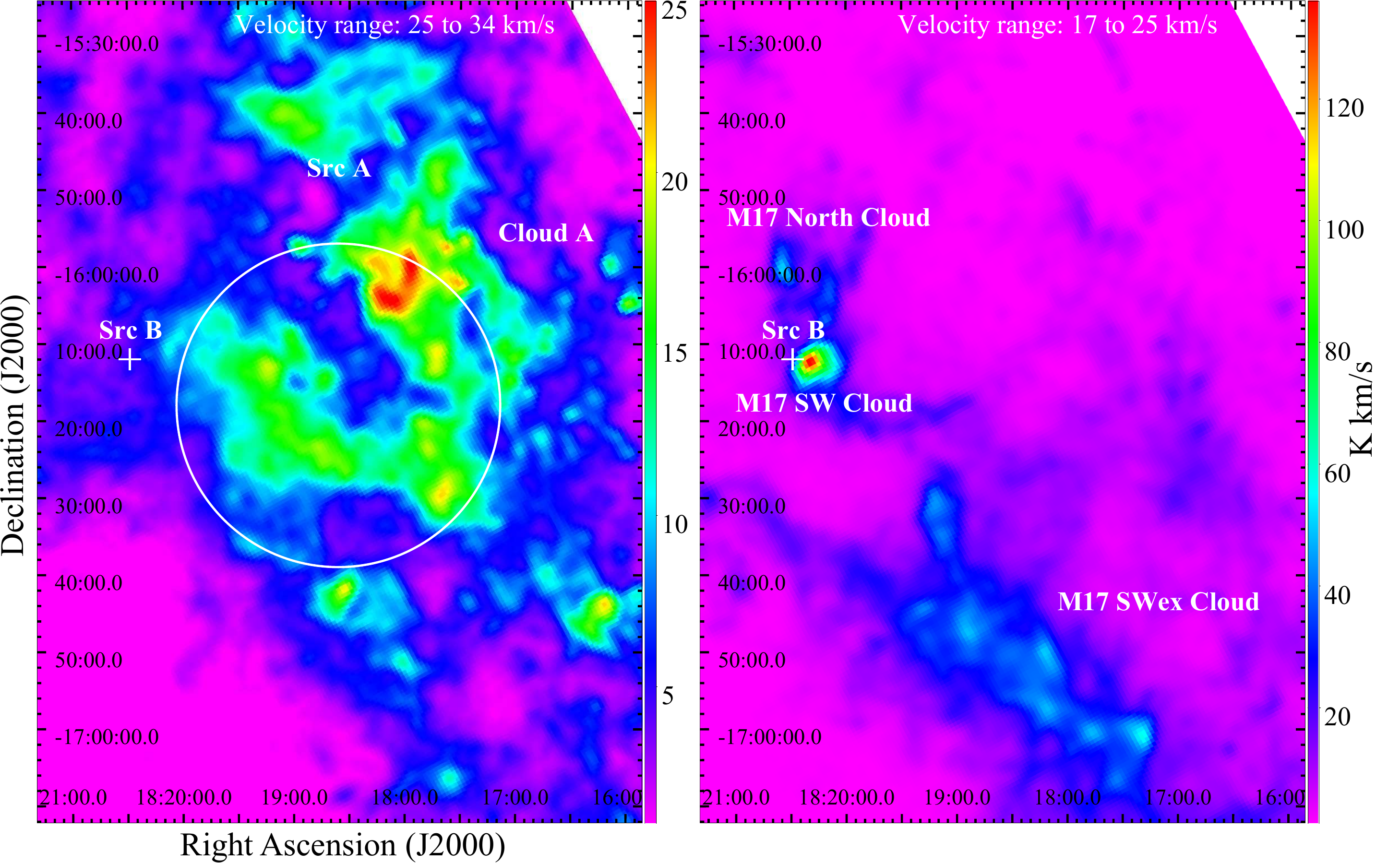}
    \caption {The integrated intensity map of $^{13}$CO ($J$=1-0) line emission in the velocity range 25-33\,$\mathrm{km\,s^{-1}}$ (left) and 17-25\,$\mathrm{km\,s^{-1}}$ (right) respectively. The white circle represents  the extended gamma-ray component (Src A), and  the white cross shows the location of  point-like component (Src B). See Sect.~\ref{sec:gas} for details.
    }
\label{fig:Src AB_CO}
\end{figure*}

\begin{figure}
    \centering
    \includegraphics[width=0.43\textwidth]{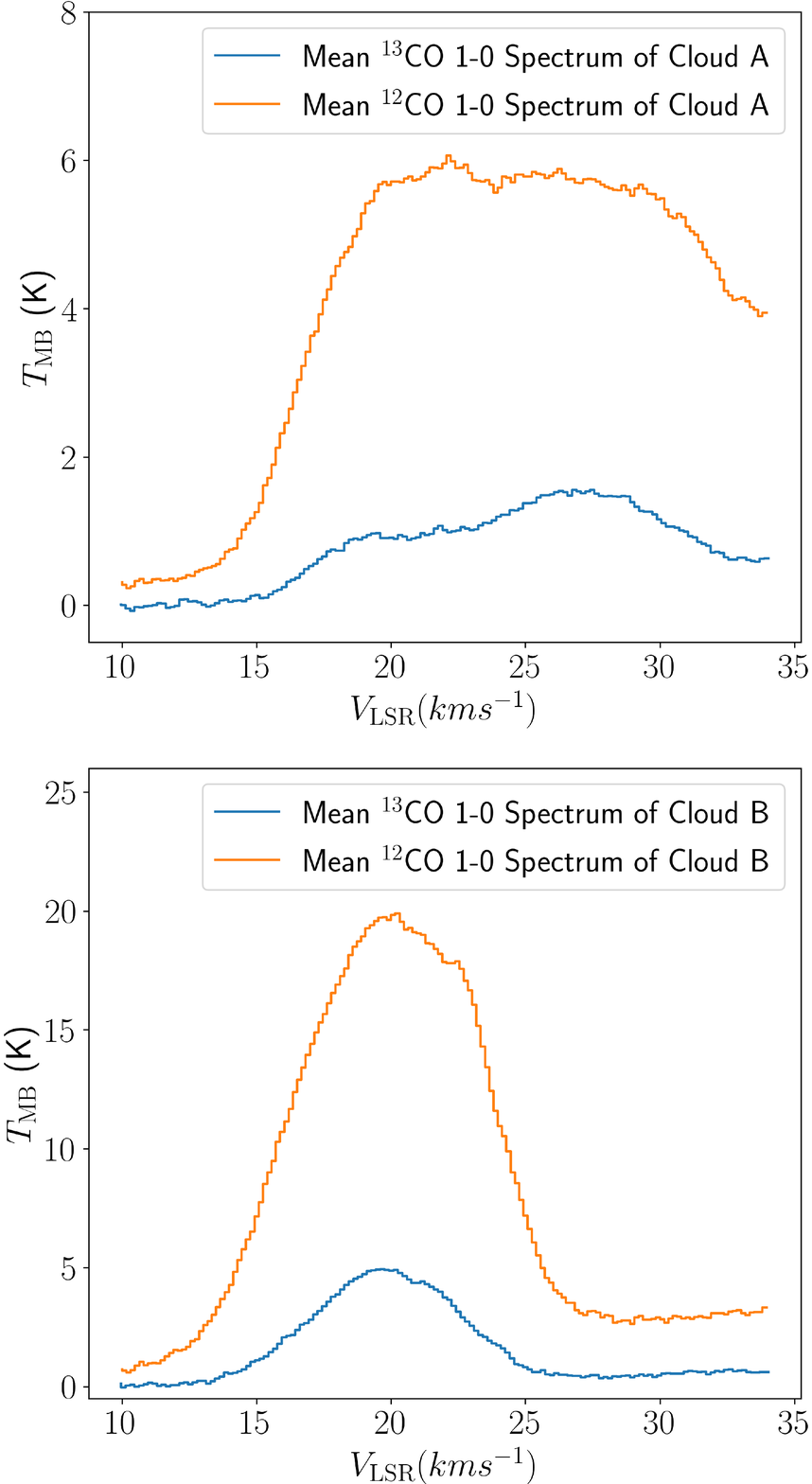}
    \caption{Mean CO spectra for molecular gases (clouds A and B) that are possibly associated with Src A (top) and Src B (bottom).}
    \label{fig:co_spectrum}
\end{figure}

As shown in Fig.\ref{fig:Src AB_CO}, Src B is much closer to the M17 SW cloud (hereafter referred to as cloud B) than to the M17 north cloud. Cloud B shows a peak intensity at the velocity $V_\mathrm{LSR}\approx20\,\mathrm{km\,s^{-1}}$, and a velocity range from 10 to $28\,\mathrm{km\,s^{-1}}$ (see Fig.\ref{fig:co_spectrum}). At velocity greater than $28\,\mathrm{km\,s^{-1}}$, the CO ($J$=1-0) line emission shows an extended distribution at a level of $T_\mathrm{MB}\sim2-3\,\mathrm{K}$. . 
The molecular gas within the region of Src A (hereafter referred to as cloud A, the white circle in Fig.\ref{fig:Src AB_CO}) shows velocity in the range from 10 to $34\,\mathrm{km\,s^{-1}}$, much broader than the velocity range of cloud B (see Fig.\ref{fig:co_spectrum}). For cloud A, the peak intensity of the $^{13}$CO ($J$=1-0) line emission is at $V_\mathrm{LSR}\approx 27\,\mathrm{km\,s^{-1}}$. However, both cloud B and M17 SWex cloud show peak intensity at $V_\mathrm{LSR}\approx20\,\mathrm{km\,s^{-1}}$. The large difference in $V_\mathrm{LSR}$, together with the far distance from Src A to cloud B and M17 SWex cloud, indicate that Src A is not associated with the molecular gas at the velocity $V_\mathrm{LSR}\approx20\,\mathrm{km\,s^{-1}}$. Thus, we suggest that Src A is associated with cloud A with a peak intensity at $V_\mathrm{LSR}\approx 27\,\mathrm{km\,s^{-1}}$. 


The bulk masses of clouds A and B depend strongly on the adopted velocity range of the $^{12}$CO line emission. For cloud B that possibly associated with Src B, a velocity range $13-25\,\mathrm{km\,s^{-1}}$ is assumed for the $^{12}$CO line emission. A wider range in velocity would definitely include more molecular gas. The molecular gas mass estimated from the velocity range $13-25\,\mathrm{km\,s^{-1}}$ represents the lower limit of the mass of cloud B. We see from Figure~\ref{fig:co_spectrum} that the $^{12}$CO line spectrum averaged over Src A shows line width of $18\,\mathrm{km\,s^{-1}}$ and equal contributions from clouds A and B. Since it is impossible for us to confirm an exact range in velocity for cloud A that is likely related to Src A,  here we adopt a velocity range $25-34\,\mathrm{km\,s^{-1}}$ for cloud A. This velocity range would miss a fraction of molecular gas with velocity exceeding this range, but preclude mostly the molecular gas at velocity about $40\,\mathrm{km\,s^{-1}}$ from the Scutum arm. Therefore, the integrated intensity of $^{12}$CO line emission ($W_{\rm CO}$) over the ranges $25-34\,\mathrm{km\,s^{-1}}$ and $13-25\,\mathrm{km\,s^{-1}}$ are obtained for clouds A and B, respectively. 

Next, we used the nominal $X$-factor, $2\times10^{20}\,\mathrm{cm^{-2}}$/($\mathrm{K\,km\,s^{-1}}$),  for the conversion from  $W_\mathrm{CO}$ to the molecular hydrogen column density $N(\mathrm{H_2})$.
For cloud A, $^{13}$CO line emission shows a prominent cavity of little emission toward the center of cloud A, while $^{12}$CO line emission toward the same cavity is still significant. Thus, for cloud A,  the average $N(\mathrm{H_2})$ estimated from the $^{13}$CO line emission (hereafter referred as $N(\mathrm{H_2})_{\rm ^{13}CO}$) is $8.1\times10^{21}\,\mathrm{cm^{-2}}$, slightly lower than the average $N(\mathrm{H_2})$ estimated from $^{12}$CO line emission ($N(\mathrm{H_2})_{\rm ^{12}CO}$), which is $1.0\times10^{22}\,\mathrm{cm^{-2}}$. 
As for cloud B, the $N(\mathrm{H_2})_{\rm ^{12}CO}$ is $4.9\times10^{22}\,\mathrm{cm^{-2}}$, which should be severely underestimated because the $^{12}$CO line emission is optically thick. In comparison, the estimated  $N(\mathrm{H_2})_{\rm ^{13}CO}$ of cloud B can reach to $3.4\times10^{23}\,\mathrm{cm^{-2}}$ \footnote{This method makes use of both $^{12}$CO and $^{13}$CO line emissions and assumes the local thermal equilibrium state for the two molecules. Detailed description of this method is referred to \citet{2022arXiv220413296C} and references therein.}.

Assuming a distance of 2.0\,kpc, the radius of cloud B is $\sim 2.5$\,pc. Thus, the total gas mass of cloud B is $2.2\times10^4\,{\rm M}_{\sun}$ according to its $N(\mathrm{H_2})_{\rm ^{12}CO}$ or $1.5\times10^5\,{\rm M}_{\sun}$ according to its $N(\mathrm{H_2})_{\rm ^{13}CO}$. \citet{Quang2020} estimated a total gas mass of $1.43\times10^5\,{\rm M}_{\sun}$ for the molecular gas near to the YMC NGC\,6618, including cloud B and the less massive M17 North cloud. Therefore, we adopt a total gas mass of $1.5\times10^5\,{\rm M}_{\sun}$.  Then the corresponding hydrogen number density $n(\mathrm{H})$ of cloud B is $4.4\times10^4\,\mathrm{cm^{-3}}$, assuming a thickness of 5\,pc. Similarly, the total gas mass of cloud A is about $1.3\times10^5\,{\rm M}_{\sun}$, comparable to that of cloud B. Assuming a thickness of 5\,pc for cloud A, a value as the same as cloud B, the hydrogen number density $n(\mathrm{H})$ of cloud A is estimated to be $1.3\times10^3\,\mathrm{cm^{-3}}$. However, if the thickness of cloud A is comparable to the projected size on the plane of sky, i.e., be as large as 25\,pc, it will yield a lower density, $\sim 2.6\times10^2\,\mathrm{cm^{-3}}$.  Thus, we suggest that $n(\mathrm{H})$ of cloud A is likely between $\sim 2.6\times10^2\,\mathrm{cm^{-3}}$ and $\sim1.3\times10^3\,\mathrm{cm^{-3}}$.  

In short, the molecular gas components possibly associated with Src A and Src B, clouds A and B, are very different in density, size, and kinematics. At a $V_\mathrm{LSR}\sim27\,\mathrm{km\,s^{-1}}$, cloud A is a very extended molecular cloud of moderate density. Cloud B is a compact dense molecular cloud with a $V_\mathrm{LSR}\sim20\,\mathrm{km\,s^{-1}}$.



\section{Discussion}
\label{sec:dis}
To investigate the possible radiation mechanisms of the gamma-rays in M17 region, we fit the SEDs of Src A and Src B with both leptonic scenario, i.e., the inverse Compton scattering (hereafter referred as IC model) and hadronic scenario, i.e., proton-proton inelastic collision (hereafter referred as PP model). The fitting was performed using Naima package\footnote{\url{http://naima.readthedocs.org/en/latest/index.html}} \citep{naima}, which includes tools to perform Markov Chain Monte Carlo (MCMC) fitting of nonthermal radiative processes to the data and allows us to implement different functions. Here  the distribution function of the parent particles was assumed to be simple power law.  The best-fit results are illustrated in Fig.\ref{fig:sedfit}, in which the black lines are the results of PP model fitting and the red lines are results of IC model fitting. The corresponding parameters are listed in Table.\ref{tab:sedab}.

Around M17 region, there are several known pulsars and SNRs (see Fig.\ref{fig:tsmaps}), but none of them overlaps with Src A or Src B. Although we cannot formally rule out the possibility that the emissions are related with some unknown pulsars or SNRs, the extension of Src A and the hard spectrum of Src B can exclude the possibilities that those sources are pulsars. However, the pulsar wind nebula related with the unknown pulsars can be a possible explanation. In addition, considering the spatial correlation of both Src A and Src B with dense gas, it is possible that the CRs escaped from SNRs in the vicinity interacting with dense gas produced the detected gamma-rays. 

Another scenario is associating  Src A and Src B with the YMC NGC\,6618. NGC\,6618 contains more than 40 OB stars \citep{hoffmeister08}, the total wind power of these stars can be estimated using the estimations in \citet{santamar06}, which amounts to more than $5\times 10^{37} {\rm erg\,s^{-1} }$. Taking into account the age of 500 kyr, the total energy injected by NGC 6618 is about $10^{51}$\,erg, which is comparable to a typical supernova explosion. Thus, this system is powerful enough to accelerate enough CRs to account for the detected gamma-ray emissions. Furthermore, the hard spectrum of Src B is also similar to other YMC systems. The soft spectrum of Src A is unique compared with other YMCs. Generally speaking, if we assume the CRs are injected continuously by the YMC, the CR energy spectrum in the vicinity can be expressed as $F\left(E\right) \sim \frac{Q\left(E\right)}{D\left(E\right)}$, where $Q\left(E\right)$ and $D\left(E\right)$ is the injection spectrum and diffusion coefficient, respectively \citep{Aharonian96aa}. Therefore, a natural explanation for the different spectral indices of Src A and Src B is a different $D\left(E\right)$ in the corresponding regions. Indeed, due to the higher density and stronger magnetic field inside the YMCs, the environment can be significantly different from the interstellar medium (ISM) in the Galaxy. The MHD turbulent cascade in the ISM can be damped out effectively and CRs will stream along field lines and transport via a process of field line random walk. Such process have been studied in detail in starburst galaxies in \citet{Krumholz20}, and the environments in YMCs are comparable to these starburst galaxies. In this case, the effective diffusion coefficient is energy independent at lower energy (below some critical energy $E_{\rm br}$), thus the propagated CR spectra are the same as the injected spectra. This provide an natural explanation for hard spectrum in Src B and in other YMC systems.  If we assume a continuous injection, the CR energy density scales as $\frac{1}{r}$, where $r$ is the distance to the CR source. Thus, the CR density can be much smaller in Src A than in Src B, if we assume the CRs are injected from NGC\,6618.  For Src A, the CR transport may be still dominated by the CR scattering with the MHD turbulence in the ISM. In this case, $D\left(E\right)$ scales as $E^{0.33...0.5}$, which predicts a softer spectrum. The difference of indices of Src A and Src B is about 0.3, which is also consistent with the index of $D\left(E\right)$ in the ISM. 

In such a scenario all the gamma-ray emissions surrounding YMCs should contain two regions, in one of which the CR transport is dominated by streaming along field line and reveal a harder spectrum (inner region); in another the CR transport is dominated by scattering with MHD turbulence in the  ISM (outer region).  For other YMC systems, the "outer region" has not been detected yet, one reason may be that for those more powerful systems such as Cygnus cocoon, Westerlund 2, and NGC\,3603, due to the larger wind power and longer age, the CR density therein is significantly higher, thus the "inner region" in which CRs transport by streaming along field lines is much larger than that in NGC\,6618, and occupy the whole dense regions near the YMCs, thus dominate the produced gamma-ray emissions. In weaker systems such as W40 and W43, the "outer region" in which CR transport dominated by diffusion in MHD turbulence in ISM are too weak to be distinguished from the diffuse gamma-ray background. 

In the case of "inner region", the CR streaming velocity is dominated by the balance between the damping rate and the CR streaming instability, which results a energy dependent diffusion at higher energy and predict a energy break of CR spectrum as well as the corresponding gamma-ray spectrum.  As estimated in \citet{Krumholz20}  the break energy scale as $E_{\rm br} \sim \left(\frac{\epsilon}{n^{1.5}}\right)^{\frac{1}{\gamma-1}}$, where $\epsilon$ is the CR energy density, $n$ is the ambient gas density and $\gamma$ is the spectral index of the CR (assuming a power law distribution). Thus, $E_{\rm br}$ can be different in different environment. Dramatically, as we mentioned above, in the case of continuous injection the CR energy density scale as $\frac{1}{r}$, which predicted a decrease of $E_{\rm br}$ with increasing $r$. For NGC\,6618, such a break are not observed. But for Cygnus cocoon, a spectral softening at several TeVs have already been detected by HAWC \citep{hawc_cygnus}, if such a softening is a consequence of the propagation effects we discussed here, a spatial dependence of the energy break is expected, which may be tested by later observations.

\begin{table} 
	\caption{ SED fit results for different radiation models of Src A and Src B}
	\begin{tabular}{cccc}
	\hline
	\hline
	 &Model & Index  & {$W_{\rm p}$ or $W_{\rm e}$} (erg) $^{a}$  \\
	\hline
\multirow{2}{*}{Src A} &PP& $3.13^{+0.14}_{-0.12}$& $4.67^{+0.21}_{-0.18}\times10^{47}$ \\ 
    & IC &$4.35^{+0.18}_{-0.31}$&$4.63^{+2.31}_{-2.16}\times10^{50}$  \\ 
    \hline
	\multirow{2}{*}{Src B}&PP& $\mathrm{2.50_{-0.10}^{+0.10}}$ & $\mathrm{1.83_{-0.10}^{+0.12}\times10^{45}}$ \\ 
	& IC &3.38$_{-0.16}^{+0.16}$ & $1.43_{-0.59}^{+0.92}\times10^{49}$  \\ 
	\hline
	\hline
\end{tabular}
\\
{\footnotesize  $^{a}$ The total energy of the parent protons or electrons (with energy $>1$\,GeV) for generate the gamma-ray emission, assuming the medium density $n(\mathrm{H})$ is $1.3\times10^3\,{\rm cm}^{-3}$ for Src A and $4.4\times10^4\,{\rm cm}^{-3}$ for Src B, and the distance of both Src A and Src B is ${2\,{\rm kpc}}$.}
\label{tab:sedab} 
\end{table}

\section{Conclusions}
\label{sec:con}

In this work, we report the detection of gamma-ray emission toward M17 region, which hosts one of the youngest YMCs in our Galaxy, NGC\,6618. In addition to Cygnus cocoon \citep{Ackermann11, aharonian19}, NGC 3603 \cite{yang17}, W40 \citep{sun20a}, and RSGC 1 \citep{sun20b}, this detection provides another case in the population of gamma-ray emitting YMCs. 
However, the gamma-ray emission of this source has its unique characteristic: the gamma-rays can be separated into two components, one is point-like and with harder spectrum and another is extended and with softer spectrum. Both components show spatial correlation with dense gases. 
Assuming the gamma-rays are produced by the interactions of CRs injected by NGC\,6618 with the ambient gas, the different spectral features can be addressed by the propagation effects investigated in  the  environments of starburst galaxies \citep{Krumholz20}. In this scenario, the MHD turbulent in the ISM can be damped out effectively, and CRs will stream along field lines and transport via a process of field line random walk. As a result, the CR diffusion coefficient is energy-independent at low energy and will become energy-dependent above certain break energy. The high densities of both gas and CRs in YMCs are similar to those in starburst galaxies, thus such a comparison is feasible. In M17 region, the hard spectrum of the point-like source Src B can be explained by the energy-independent diffusion. As for Src A, due to the lower gas density and CR density, the damping mechanism can be switched off, meanwhile, the diffusion coefficient are still the same as in the ISM. The features related to these propagation effects have not been observed in other YMC systems yet. But the observed energy break in Cygnus cocoon region can be explained naturally in such a scenario. Since YMCs can potentially be an alternative CR source, the propagation of CRs in the vicinity of these sources would be crucial to understand the injection of CRs to the ISM, which can be revealed by further observations with more spatial and spectral information. 

\section{Acknowledgements}
This research made use of the data from the Milky Way Imaging Scroll Painting (MWISP) project, which is a multi-line survey in $^{12}$CO/$^{13}$CO/${\rm C}^{18}$O along the northern galactic plane with PMO-13.7m telescope. MWISP was sponsored by National Key R\&D Program of China with grant 2017YFA0402701 and CAS Key Research Program of Frontier Sciences with grant QYZDJ-SSW-SLH047. Bing Liu is supported by the NSFC under grant 12103049. Ruizhi Yang is supported by  the NSFC under grants 12041305, 11421303 and  the national youth thousand talents program in China. Zhiwei Chen acknowledges support from NSFC grants 11903083, 12173090, U2031202, 11873093, and 11873094.

\section{Data Availability}
The Fermi-LAT data used in this work is publicly available, which is provided online by the NASA-GSFC Fermi Science Support Center \footnote{\url{https://fermi.gsfc.nasa.gov/ssc/data/access/lat/}}. A MWISP open data proposal is required to access the CO line data used in this work,  and the form could be downloaded from \url{http://english.dlh.pmo.cas.cn/op/odp/}. 



\bibliographystyle{mnras}
\bibliography{cite_m17} 








\bsp	
\label{lastpage}
\end{document}